\documentclass[pra,preprint,amsmath,amssymb]{revtex4}
\usepackage{graphicx}
\usepackage{bm}
\usepackage{amsmath,amssymb}
\usepackage{color}

\begin{document}
\title{Measurement-based quantum computation cannot avoid byproducts}  
\author{Tomoyuki Morimae}
\email{morimae@gmail.com}
\affiliation{Controlled Quantum Dynamics Theory Group, Imperial College London, London SW7 2AZ, United Kingdom}
\date{\today}
            
\maketitle  
Measurement-based quantum computation~\cite{MBQC}
is a novel model of quantum computing
where universal quantum computation can be done
with only local measurements on each particle of 
a quantum many-body state,
which is called a resource state~\cite{MBQC,Brennen,Miyakeholo,Miyake2d,Wei2d,
Gross,stringnet,tricluster,FM_topo,Li_topo,
Cai,CV,SR}. 
One large difference of the measurement-based model from the circuit
model is the existence of ``byproducts".
In the circuit model, a desired unitary $U$ can be implemented
deterministically, whereas the measurement-based
model implements $BU$, 
where $B$ is an additional operator, which is called
a byproduct. 
In order to compensate byproducts,
following measurement angles
must be adjusted.
Such a feed-forwarding requires some classical processing and
tuning of the measurement device, which 
cause the delay of computation and the 
additional decoherence. 
Is there any ``byproduct-free" resource state?
Here we show that if we respect
the no-signaling principle~\cite{Popescu},
which is one of the most fundamental principles of physics,
no universal resource state can avoid byproducts.

\section{No-signaling principle}
The no-signaling principle means that no message
can be transmitted immediately.
It is one of the most central principles in physics,
and it is known to be more fundamental than
quantum physics in the sense that there is a no-signaling theory
which is more non-local than quantum physics~\cite{Popescu}.

\begin{figure}[htbp]
\begin{center}
\includegraphics[width=0.35\textwidth]{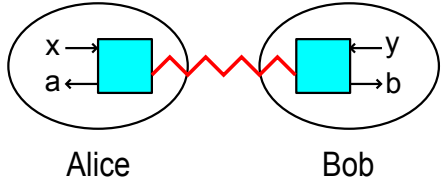}
\end{center}
\caption{
{\bf The setup for the no-signaling principle.}
Alice and Bob share a physical system.
Alice inputs $x$ and obtains the output $a$.
Bob inputs $y$ and obtains the output $b$.
} 
\label{nosignaling}
\end{figure}

The definition of the no-signaling principle is as follows~\cite{Popescu}.
Let us assume that Alice and Bob share a physical system,
which might be classical, quantum, or even super-quantum system
(Fig.~\ref{nosignaling}).
Each of them measure their own part.
Then, the no-signaling means
\begin{eqnarray*}
\sum_a P(a,b|x,y)=\sum_a P(a,b|x',y)
\end{eqnarray*}
for all $b$, $x$, $x'$, and $y$.
Here,
$a$ is Alice's measurement result,
$b$ is Bob's measurement result, $x$ represents 
Alice's measurement choice, and $y$ represents Bob's measurement choice.
Intuitively, this equation means that
even if Alice changes her measurement choice,
Bob's measurement result is not affected.

\section{Measurement-based quantum computation}
\begin{figure}[htbp]
\begin{center}
\includegraphics[width=0.2\textwidth]{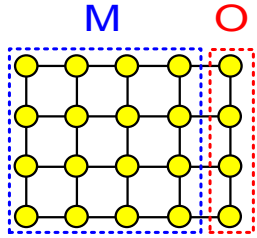}
\end{center}
\caption{
{\bf The resource state.}
The set $M$ of particles to be measured, and
the set $O$ of particles which contains the quantum output state
at the end of the computation.
} 
\label{resource}
\end{figure}

We perform measurement-based quantum computation on the quantum
many-body system $S$ whose state is $\rho$.
In other words, $\rho$ is the resource state
of our measurement-based quantum computation.
We divide $S$ into two subsystems $M$ and $O$
such that
$S=M\cup O$ and $M\cap O=\phi$.
The subsystem $M$ is the set of particles to be measured,
and the subsystem $O$ is the set of particles which 
contains the quantum output at the end of the computation (Fig.~\ref{resource}).

Let us assume that our quantum computation
is the implementation of a unitary operator $U$ on 
the initial state $\sigma$.
For that purpose, we measure all particles in $M$.
After the measurement, 
the state of the subsystem $O$ becomes
\begin{eqnarray*}
B_kU\sigma U^\dagger B_k^\dagger
\end{eqnarray*}
with probability $p_k$,
where $k$ is a vector which represents specific measurement
results (a branch),
and $B_k$ is the byproduct
corresponding to $k$.
If $B_k=I$ for all $k$, this means that
the resource state $\rho$  
is ``byproduct free" for $U$. 

For example, let us consider the one-dimensional cluster state~\cite{MBQC}.
If we measure a qubit of the cluster state, 
we can implement $X^sHe^{iZ\theta/2}$
if the measurement result is $s=0,1$, respectively.
These two measurement outcomes occur with the equal probability.
Therefore, in the cluster model, 
$X^pZ^qU$, where $p,q\in\{0,1\}$, is implemented with equal
probabilities $1/4$.
Although such a byproduct $X^pZ^q$ plays a crucial role
in some secure computation, such as blind quantum 
computation~\cite{BFK,FK,Barz,Vedran,AKLTblind,topologicalblind,MABQC,CVblind,
topoveri},
it would be desirable for a scalable quantum computation
if we could find a resource state which
is byproduct free, since byproducts require feed-forwarding,
and it causes an additional decoherence.
However,
we show that due to the no-signaling principle,
no universal resource state
is byproduct free.

\section{Proof}
\begin{figure}[htbp]
\begin{center}
\includegraphics[width=0.4\textwidth]{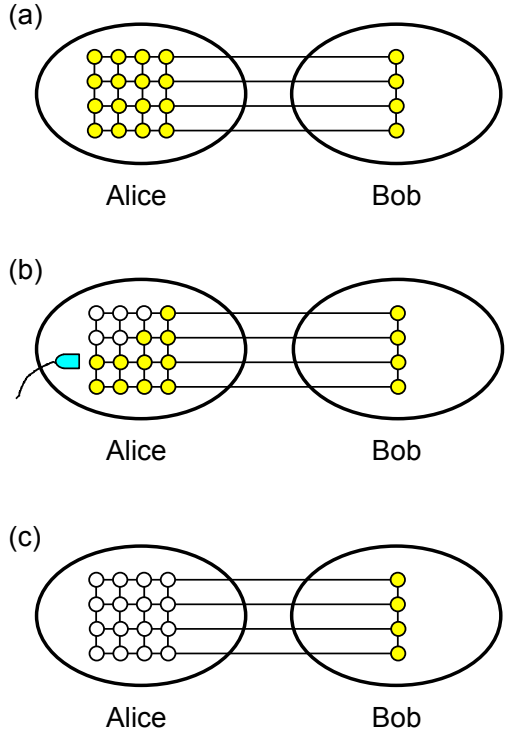}
\end{center}
\caption{
{\bf Measurement-based quantum computation between Alice and Bob.}
(a) Alice and Bob share the resource state.
(b) Alice measures each particle according to her algorithm.
(c) Bob has the output state.
} 
\label{AliceandBob}
\end{figure}

In order to show it,
let us consider measurement-based quantum computation 
between
two people, Alice and Bob, as is shown in
Fig.~\ref{AliceandBob}. 
Alice and Bob
first share the resource state $\rho$
(Fig.~\ref{AliceandBob} (a)).
Alice possesses the subsystem $M$ and Bob possesses the subsystem $O$.
Next Alice measures each particle of $M$ 
(Fig.~\ref{AliceandBob} (b)).
After Alice measuring all particles of $M$, the subsystem $O$,
which Bob has, is in the state
\begin{eqnarray*}
B_kU\sigma U^\dagger B_k^\dagger
\end{eqnarray*}
with the probability $p_k$
from Alice's view point
(Fig.~\ref{AliceandBob} (c)).
From Bob's view point, the state of the subsystem $O$ is
\begin{eqnarray*}
\sum_kp_k
B_kU\sigma U^\dagger B_k^\dagger,
\end{eqnarray*}
since Bob does not know $k$.
Let us assume $B_k=I$ for all $k$. 
Then, Bob's state is
\begin{eqnarray*}
\sum_kp_k B_kU\sigma U^\dagger B_k^\dagger
&=&
\sum_kp_kU\sigma U^\dagger\\
&=&
U\sigma U^\dagger.
\end{eqnarray*}

Let us choose
a unitary operator $V$ such that
\begin{eqnarray*}
V\sigma V^\dagger\neq U\sigma U^\dagger.
\end{eqnarray*}
Because $\rho$ is a universal resource state, the implementation
of $V$ is also possible with appropriate measurements on $M$.
Then, if we repeat the above argument by
replacing $U$ with $V$,
what Bob finally obtains after the two-party measurement-based
quantum computation (Fig.~\ref{AliceandBob} (c)) is
\begin{eqnarray*}
V\sigma V^\dagger
\end{eqnarray*}
if $\rho$ is byproduct free for $V$.
However, since we have assumed 
\begin{eqnarray*}
V\sigma V^\dagger\neq U\sigma U^\dagger,
\end{eqnarray*}
it contradicts to the no-signaling principle.
In other words, 
if there exists a byproduct-free universal resource state,
Alice can have Bob possess two different states,
$U\sigma U^\dagger$ or $V\sigma V^\dagger$, 
at her will
by choosing her measurement pattern on $M$.
It contradicts to the no-signaling principle,
since she can exploit it to transmit her message to Bob.

\section{Discussion}
In this paper,
we have shown that if we respect the no-signaling principle,
no universal resource state of measurement-based quantum
computation can be byproduct free.

In the above argument,
we have considered the deterministic measurement-based quantum computation.
In other words, we have assumed that a desired unitary operator
can be implemented up to byproducts by measuring a {\it fixed} number of
particles.
For example, the cluster state allows such a deterministic
measurement-based quantum computation.
However, there are some resource states which allow only
non-deterministic measurement-based quantum computation.
For example, let us consider the one-dimensional
Affleck-Kennedy-Lieb-Tasaki (AKLT) state~\cite{AKLT,Brennen},
which is a chain of spin-1 (qutrit) particles.
If we measure a qutrit, we can implement
$Xe^{iZ\theta/2}$, $XZe^{iZ\theta/2}$, or $Z$
according to the measurement result.
Note that if we obtain the third measurement result, we cannot
implement the $z$-rotation,
and we just obtain the trivial $Z$ operation.
Hence a desired unitary operator 
can be implemented up to byproducts with the probability
$1-3^{-r}$ if we measure $r$ qutrits.

In this case, a similar argument leads to
\begin{eqnarray*}
(1-3^{-r})U\sigma U^\dagger+3^{-r}F_U\sigma F_U^\dagger
=
(1-3^{-r})V\sigma V^\dagger+3^{-r}F_V\sigma F_V^\dagger,
\end{eqnarray*}
where $F_U$ and $F_V$ are some ``failed" operators,
which are bounded.
Then,
we obtain
\begin{eqnarray*}
\|U\sigma U^\dagger-V\sigma V^\dagger\|
&=&\frac{3^{-r}}{1-3^{-r}}\|F_V\sigma F_V^\dagger-F_U\sigma F_U^\dagger\|\\
&\le&\frac{3^{-r}}{1-3^{-r}}\Big(
\|F_V\sigma F_V^\dagger\|+\|F_U\sigma F_U^\dagger\|\Big),
\end{eqnarray*}
which again leads to the contradiction if we choose
$V$ in such a way that it is sufficiently different from $U$.

\acknowledgements
The author acknowledges JSPS for support.


\end{document}